\documentclass[a4paper,11pt]{article}
\usepackage{jinstpub} 
\usepackage{lineno}
\usepackage{natbib}
\usepackage{tabularx}
\usepackage{booktabs}

\title{\boldmath Geant4 and FLUKA Simulations of a Cyclotron Based 30 MeV Proton-Beryllium Reaction: Benchmarking and Optimization of Neutron Fields
}

 \author[a,1]{E. Gover\note{Corresponding author.}}
 \author[a,b,c]{D. Veske}
 \author[a]{M.B. Demirkoz}
 \affiliation[a]{Department of Physics, Middle East Technical University, \c{C}ankaya/Ankara, 06800, Turkey}
 \affiliation[b]{Uzay ve H{\i}zland{\i}r{\i}c{\i} Teknolojileri Uygulama ve Ara\c{s}t{\i}rma Merkezi, Orta Do\u{g}u Teknik \"Universitesi, \c{C}ankaya, Ankara 06800, T\"urkiye}
 \affiliation[c]{Columbia Astrophysics Laboratory, Columbia University in the City of New York, New York, NY 10027, USA}

\emailAdd{egemen.gover@metu.edu.tr, veske@metu.edu.tr, demirkoz@metu.edu.tr}

\abstract{For studies where a reliable neutron source/beam is required and a nuclear reactor is not a viable option (considering their high neutron flux supply, which may not be suitable for research concerning low flux operations), alternative approaches may be sought. We present a comprehensive simulation analysis of 30 MeV proton induced ${}^9\mathrm{Be}\text{(p,n)}{}^9\mathrm{B}$ reaction, which can be utilized as an isotropic neutron source. This energy was chosen as it was the maximum allowed limit of protons supplied by IBA's Cyclone 30 XP proton cyclotron.  Due to different underlying physics and transport mechanisms, when it comes to numerical calculations, slight variations in findings between different toolkits may occur. Hence it may require one to have a guide at hand to address the differences and interpret the data more accurately before conducting the actual experiment. In this work present the different numerical results of Geant4 and FLUKA. We conducted preliminary Monte Carlo runs to estimate the resulting neutron fluence and prompt gamma dose equivalents, while also assessing the degree of moderation achieved by different thicknesses of high density polyethylene. Furthermore, this work also presents an examplar modular irradiation station designed for target-moderator configurations, with the capability of generating thermal neutron fields.}


\arxivnumber{0} 

\begin{document}
\maketitle
\flushbottom

\section{Introduction}
\label{sec:intro}

Neutron beam applications spans a wide area of research. Material hardness studies, clinical neutron radiography, neutron diffraction studies, investigation of neutron induced activation reactions, neutron activation analysis serve as exemplars of the full spectrum of outcomes attainable through a reliable neutron beam source. Moreover, SEU (single event upset) tests for electronic boards that will be subject to neutron radiation in upper atmosphere where the dominant source of upsets are neutrons can also be conducted with a novel neutron source.The growing demand for neutrons across various research fields has necessitated the exploration of alternative neutron sources.  For facilities operating a proton- or deuteron-based accelerators, one method is utilizing the ${}^9\mathrm{Be}\text{(p,n)}{}^9\mathrm{B}$ reaction branch to construct and orient a neutron beam. For the purpose of optimizing the irradiation parameters and benchmarking prior to tests, simulation codes such as GEANT4~\cite{geant1,geant2,geant3} and FLUKA~\cite{fluka1,fluka2} shall be deployed. Although both simulation toolkits offer solid  and consistent numerical groundwork, due to their dissimilarities in underlying physics models and code structure, there can be slight variations in computational output. This study will examine and compare the variations observed between the toolkits. To allow for an unbiased comparison, all simulation parameters will be kept the same. 

\section{Optimal Target Geometry and Orientation}

In Figure \ref{fig:outergeom}, \ref{fig:innergeom} and \ref{3d_setup} , a 3D section view of the proposed experiment is shown. Target composition is 100\% pure beryllium. The isotropic fraction of beryllium is also assumed to be 100\% ${}^9\mathrm{Be}$. The choice of target was decided according to gamma/neutron ratio and thermal conductivity of preferred target material. ${}^{7}\text{Li}$ yields better neutron production results~\cite{lone1977, graves1970} but has lower thermal conductivity and higher gamma yield. Therefore, to utilize the present active water cooling effectively beryllium was chosen as target material. With respect to beam direction, the target is rotated with 45$^\circ$ angle. This was found to be the optimal positioning after various runs with different angles. This results in slight diminishment in back scattering of neutrons. resulting neutron fluence was registered just before the moderation block, where only neutron steps that enter the boundary were counted.  Immediately downstream of the target, a 20 mm thick aluminum beam dump is strategically positioned. This component fulfills two critical functions. It effectively intercepts and absorbs the incident proton beam, halting its trajectory, and the target material exhibits a thermal conductivity of 210 W/mK~\cite{jeon2020target}, rendering it amenable to efficient water cooling. The aluminum beam dump facilitates this cooling process by acting as a heat sink, dissipating the energy deposited in the target to prevent melting (beryllium is highly corrosive and toxic).  This implies proton-aluminum reaction will also contribute to neutron fluence. For attenuation of gamma rays, 5 cm thick lead shell encapsulates the whole setup. To accurately simulate the radiation environment to correctly assess neutron fluence and radiation dose within a cyclotron research and development setting, the construction of the irradiation chamber necessitates the incorporation of substantial shielding. In alignment with the design principles observed at the TENMAK-NUKEN Proton Accelerator Facility in Turkey, a concrete wall thickness of 1.5 meters is deemed appropriate for this purpose. 

Table \ref{tab:target_angle} presents neutron fluences corresponding to various target angular orientations. As $45^\circ$ angle results in the highest yield, it was chosen for target orientation. 
\begin{table}[htbp]
\centering
\caption[Total fluence response (particles per area per primary) due to different target tilts at a point \(P(x,0,0)\) on the \(x\)-axis.]{Total fluence response (particles per area per primary) due to different target tilts at a point \(P(x,0,0)\) on the \(x\)-axis for 30 MeV proton. Obtained with FLUKA using USRBDX.}\label{tab:target_angle}
\begin{tabularx}{0.6\textwidth}{@{} 
    >{\centering\arraybackslash}X 
    >{\centering\arraybackslash}X 
  @{}}
\toprule
\textbf{Angle (°)} 
  & \textbf{Total Response (neutron/cm$^2$/proton)} \\
\midrule
\(45\) 
  & \(3.3\times10^{-5}\ \pm 0.6\% \) \\
\(30\) 
  & \(3.0\times10^{-5} \pm 3.0\%\) \\
\(20\) 
  & \(2.8\times10^{-5} \pm 1.1\%\) \\
\(10\) 
  & \(2.6\times10^{-5} \pm 2.3\%\) \\
\(0\) (perpendicular) 
  & \(2.2\times10^{-5} \pm 0.8\%\) \\
\bottomrule
\end{tabularx}
\end{table}

After determining the optimal target angle, we then moved forward to find the optimal incident energy. To do so, spherical neutron counts were obtained for different energies. Using Geant4, the following trend was acquired.

\begin{figure}[h]
    \centering
    \begin{minipage}{0.48\textwidth}
        \centering
        \includegraphics[width=\linewidth]{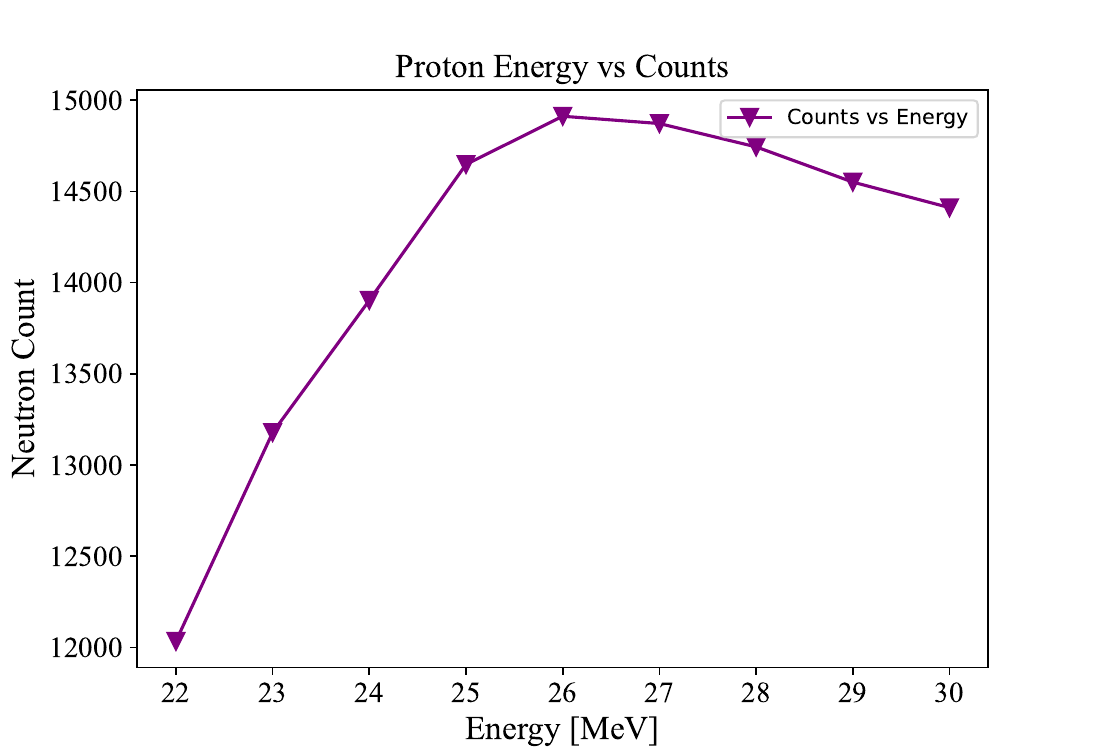}

    \end{minipage}
    \hfill
    \begin{minipage}{0.49\textwidth}
        \centering
        \includegraphics[width=\linewidth]{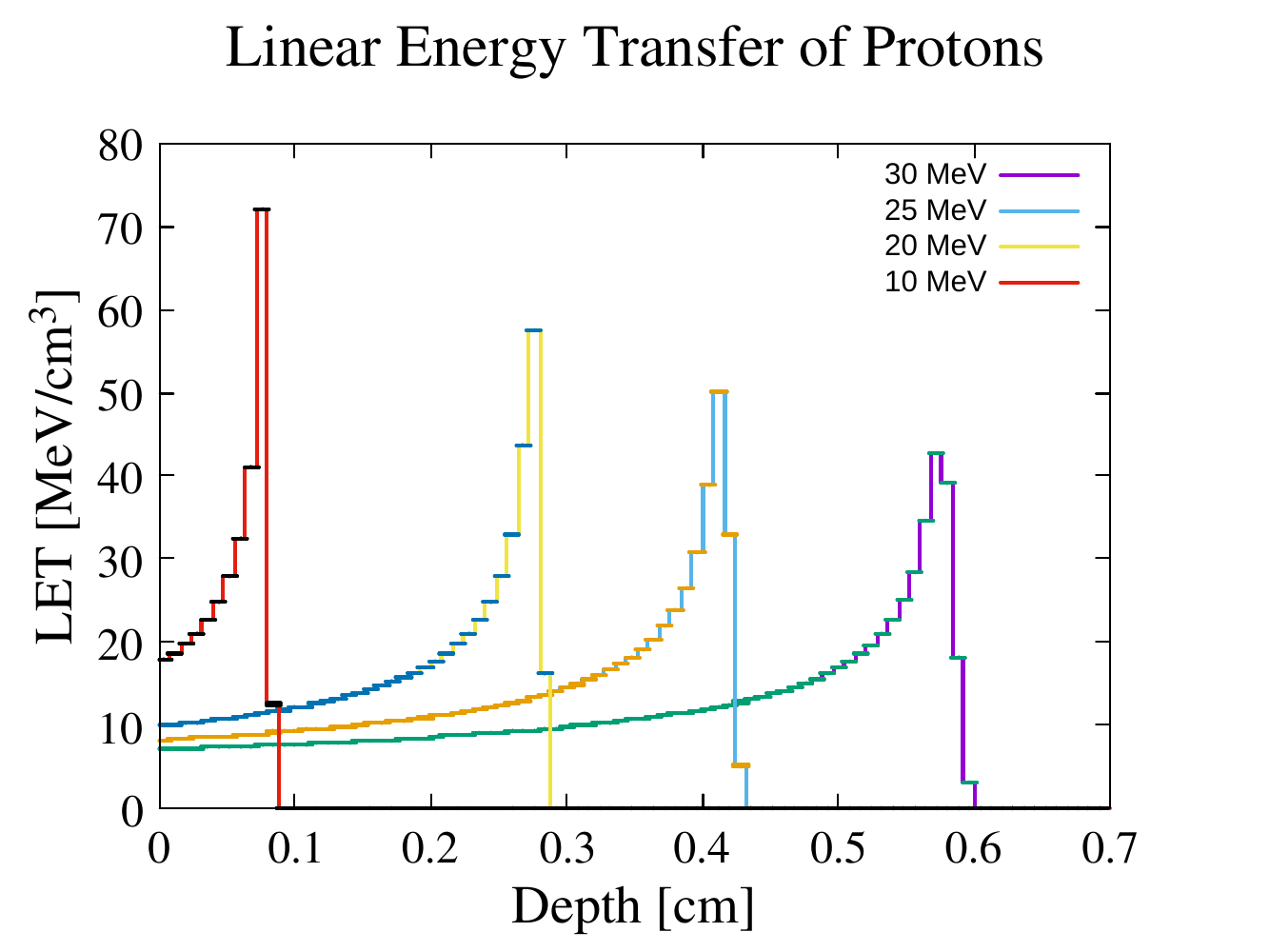}
    \end{minipage} \caption{Different incident proton energies were tested to get the neutron count coming out of an inclined target (left). Penetration depth of protons with respective energies (right). \textit{Geant4} and \textit{FLUKA}}\label{fig:proton_penetration}
\end{figure}
In Figure \ref{fig:proton_penetration}, a perpendicular target with 0$^\circ$ tilt was tested. It suggests the highest neutron production occurs at 26 MeV, but for a tilted target the effective path length of protons increase. Higher path lenth favors higher energy protons.
\begin{table}[h]
\centering
\caption[Total neutron counts for different beryllium thicknesses (30 MeV proton, \(45^\circ\) target angle).]{Total neutron counts for different beryllium thicknesses (30 MeV proton, \(45^\circ\) target angle). Obtained with Geant4.}
\label{tab:neutroncount}
\begin{tabularx}{0.5\textwidth}{@{} 
    >{\centering\arraybackslash}X 
    >{\centering\arraybackslash}X 
  @{}}
\toprule
\textbf{Thickness (mm)} 
  & \textbf{Counts} \\
\midrule
2.0 & 14 441 \(\pm 120\) \\
3.0 & 21 244 \(\pm 146\)\\
4.0 & 24 074 \(\pm 155\)\\
5.0 & 24 260 \(\pm 156\)\\
\bottomrule
\end{tabularx}
\end{table}

Table \ref{tab:neutroncount} shows the total spherical neutron count with respect to different target thicknesses. To overcome the blistering effect of protons, the chosen target thickness should be thinner than the stopping range of protons with 30 MeV energy.

Figure \ref{fig:proton_penetration}  (right) suggests that in order to not deal with blistering of hydrogen, proton's range of 5.7 mm should be taken into consideration. The target thickness shall not exceed this value, otherwise gradual buildup of hydrogen gas inside the target volume will lead to brittlement. It is imperative to account for the target tilt in proton range calculations. A tilted target increases the effective path length of protons. For a 3 mm thick target:
\begin{equation}\label{eq:cos45}
    t_\text{eff} = \frac{t}{\cos{45^\circ}} \approx 1.414 \; t = 4.242~{\rm mm}
\end{equation}
which falls behind of 5.7 mm. If the target were to be 4 mm:
\begin{equation}\label{}
  t_\text{eff} = 5.656~{\rm mm}
\end{equation}
This value is right at the edge of blistering threshold, and hence should be avoided for safety purposes. The following results were done for $15 \ \text{mm} \times 15 \ \text{mm} \times 3 \ \text{mm}$

\section{GEANT4 Configuration and Code Description }
Geant4 (GEometry ANd Tracking 4) is an open source Monte Carlo based simulation toolkit designed and developed for particle interactions and transportations. It was developed by Geant4 Collaboration~\cite{geant1,geant2} which is an international community of scientists actively improving and maintaining the toolkit. The toolkit emphasizes object-oriented programming by utilizing C++.  It allows flexible geometry and physics registration. Compared with FLUKA, it is more configurable in the sense that the user can switch between different models such as Binary Cascade, Bertini Cascade, INCL++ etc. For our work, suitable physics lists and geometries were presented to optimize the numerical output data which better simulates neutron production channels and the neutron transportation and interaction in medium.
\begin{figure}[h]
    \centering
    \begin{minipage}{0.48\textwidth} 
        \centering
        \includegraphics[width=\linewidth]{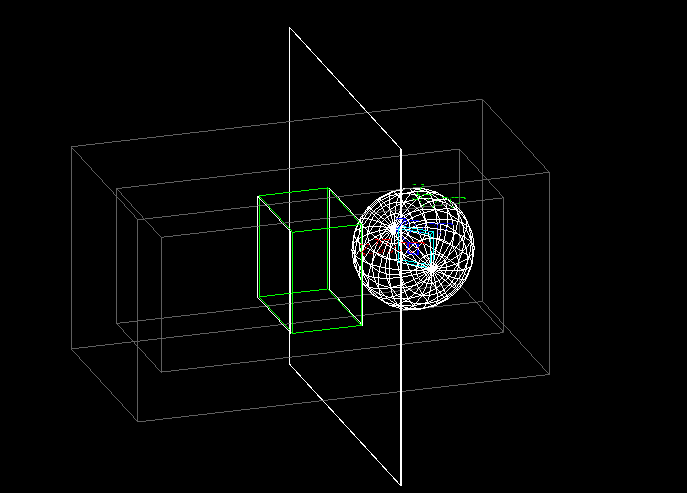}
        \caption{Outer geometry showing moderator(green), lead casing and target-beam dump duo.}
        \label{fig:outergeom}
    \end{minipage}
    \hfill
    \begin{minipage}{0.48\textwidth}
        \centering
        \includegraphics[width=\linewidth]{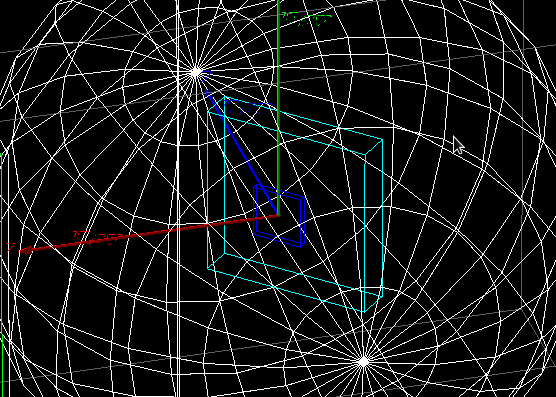}
        \caption{Inner geometry showing beam dump (light blue) and target(dark blue).}
        \label{fig:innergeom}
    \end{minipage}
\end{figure}

In Figure \ref{fig:outergeom} and \ref{fig:innergeom}, Geant4 geometry of proposed target moderator system with lead encapsulation for gamma shielding is shown.  A Gaussian proton beam with $\sigma$ of $500  
 \text{ keV/c}$ and diameter of 1 cm traverses in Z axis (blue). In Figure \ref{fig:outergeom}, a rectangular moderator (green) with a plane sheet in front can be seen. The thin plane sheet was used for neutron fluence projection on YZ plane to conduct neutron beam analysis. The beryllium target  has dimensions 15 mm x 15 mm x 3 mm and rests on an aluminum beam dump with 45$^\circ$ inclination. A thin spherical shell encompassing the beam dump-target system is implemented to measure the angular distribution of quasi-isotropic neutrons with respect to polar and azimuthal coordinates. In both simulation toolkits, target is centered at origin (0,0,0).

\subsection{Geant4 Physics}\label{physics_list}

In this work, slight modifications were imposed on neutron source example that is preconfigured and present in GEANT4 directory. As both this example and our goal handles hadronic cascade reactions, choice of hadron inelastic physics plays an important role as different physics lists can yield different results. For that, QSP\_BIC\_HP physics list for inelastic and HadronElasticPhysicsHP for elastic scattering have been registered. High precision (HP) addition carries utmost importance as the resulting neutron energy ranges from thermal to fast. HP enables the transport of thermal neutrons (or more generally, neutron energies < 19.5 MeV). G4NDL 3.16 (Neutron Data Library) was set for ENDF/B-VII to retrieve thermal neutron cross section. The complete physics list is given in Table \ref{tab:physics_summary}.
\begin{table}[htbp]
\centering
\small
\begin{tabularx}{\textwidth}{lX}
\toprule
\textbf{Geant4 Physics Constructor} & \textbf{Process Description} \\ \midrule
\texttt{HadronElasticPhysicsHP}      & High-precision elastic scattering, optimized for low-energy neutrons. \\
\texttt{G4HadronPhysicsQGSP\_BIC\_HP} & Inelastic hadron scattering via QGSP (High-E) and Binary Cascade (Int-E). \\
\texttt{G4IonElasticPhysics}         & Elastic scattering processes for alpha particles and heavier nuclei. \\
\texttt{G4IonPhysicsXS}              & Inelastic ion interactions, including spallation and fragmentation. \\
\texttt{GammaNuclearPhysics}         & Photonuclear reactions and gamma-induced nuclear disintegration. \\
\texttt{ElectromagneticPhysics}      & Standard EM interactions: ionization, Bremsstrahlung, and Compton scattering. \\
\texttt{G4DecayPhysics}              & Handles the decay processes for unstable particles. \\
\texttt{RadioactiveDecayPhysics}     & Nuclear $\alpha, \beta, \gamma$ emission and spontaneous fission for dose accuracy. \\ \bottomrule
\end{tabularx}
\caption{Summary of Physics Lists implemented in the simulation.}
\label{tab:physics_summary}
\end{table}

\subsection{Log Energy Binning}

To interpret and analyze the wide spectrum of neutron energy, logarithmic binning was essential. The binning in Geant4 is done manually in post calculation unlike the user routines feature in FLUKA. Results were acquired in counts, hence the redistribution of neutron counts to their corresponding energy bin intervals was done by a ROOT macro file.  The logarithmic energy binning was applied according to~\cite{logbinning}:
\begin{equation}
E_i=E_{\min } \times\left(\frac{E_{\max }}{E_{\min }}\right)^{\frac{i}{N}}
\end{equation}
This ensured both FLUKA and Geant4 gave results in the same dimensions and physical quantities.
\section{FLUKA Configuration and Code Description}
FLUKA (FLUktuierende KAskade), is a similar Monte Carlo simulation toolkit that serves the purpose of modeling particle interactions and transportations in a medium. It was a result of joint work between CERN (European Organization for Nuclear Research) and INFN (Istituto Nazionale di Fisica Nucleare) but since 2019~\cite{Ahdida:dispute}, there have been two distinct versions of FLUKA, each being maintained by CERN and INFN independently.  It uses both semi-empirical and theoretical models. Most of them are proprietary so unlike Geant4, this limits the configurability. The models are more rigid in FLUKA. This work utilizes FLUKA.CERN version.

\begin{figure}[h]
\centering
    \includegraphics[width=0.6\linewidth]{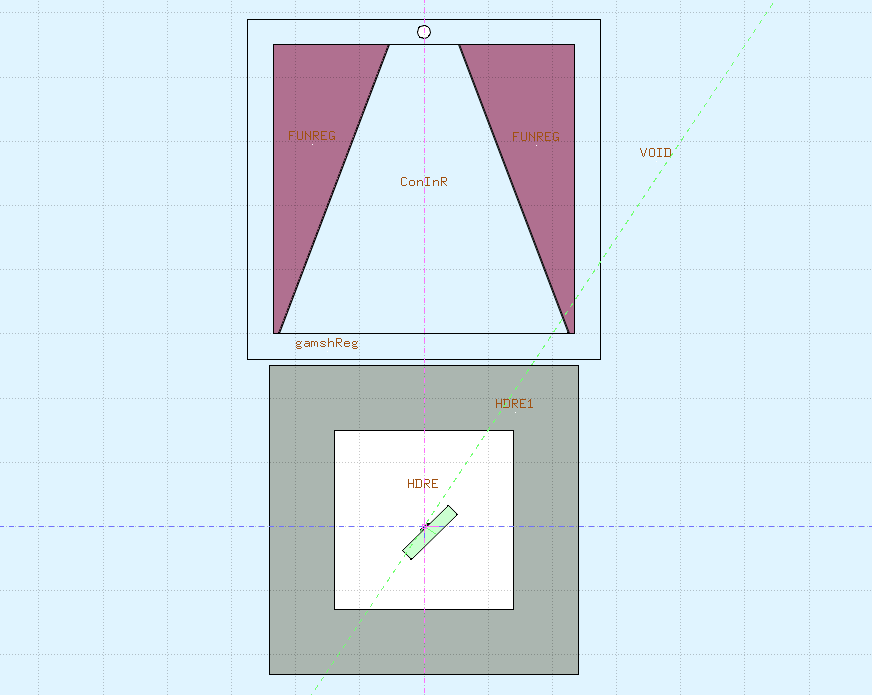}
      \caption{Top view of the constructed geometry using FLAIR and GeoViewer, graphical interface of FLUKA~\cite{flair}.\label{3d_setup}}
\end{figure}

\begin{figure}[h]
    \centering
    \includegraphics[width=0.6\linewidth]{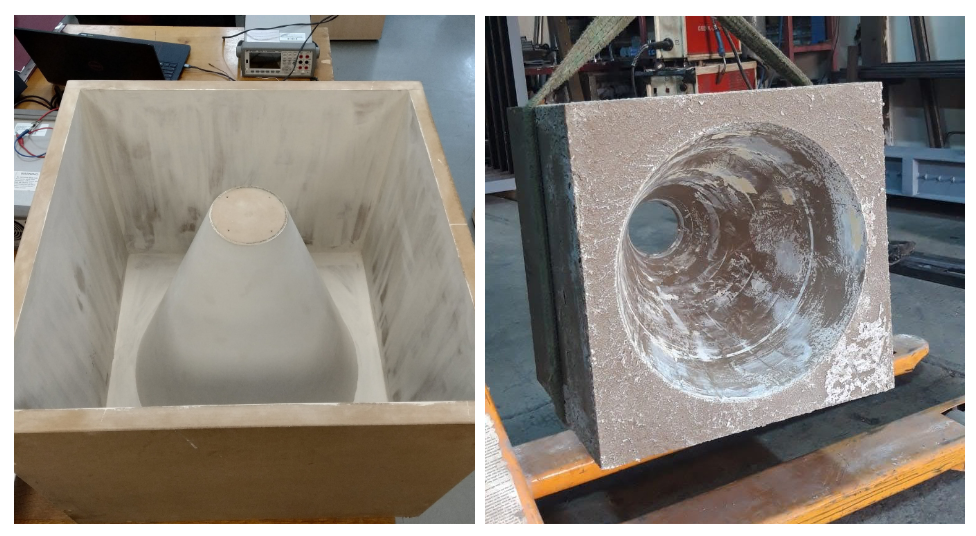}
    \caption{A neutron funnel that was used by IVMER in previous neutron studies. Left is the mold, and right is the finished product. \cite{muçogllava_2022} }
    \label{fig:funnel}
\end{figure}

In Figure \ref{3d_setup}, a top view of the simulated setup is given. Target-dump duo is surrounded by a 10 cm thick polyethylene block. It was found that encapsulating target with HDPE rather than positioning a HDPE slab on the neutron beam path results in higher neutron yield, due to the neutron reflecting properties of polyethylene. A high density concrete funnel whose  elemental composition is hydrogen (6.87 wt\%), oxygen (54.1 wt\%), carbon (27.7 wt\%), calcium (13.9 wt\%), silicon (1.38 wt\%), boron (1.27 wt\%), magnesium (0.838 wt\%), and strontium (0.512 wt\%)~\cite{muçogllava_2022}
was introduced to channel the neutrons to the desired location. The inner funnel was coated with 2.0 mm paraffin to enhance neutron moderation and scattering, while also reducing gamma-ray backscattering. A spherical probe of radius 1.0 cm was positioned at point P(70) (x = 70.0 cm) to register the neutrons. In Figure \ref{fig:funnel}, the neutron funnel is given. Exact geometry was constructed in FLUKA.

\subsection{FLUKA Physics}
All runs were done with FLUKA version 4-3.3. For base level predefined physics settings, “PRECISION” option was selected. This option ensures low energy neutrons are properly transported (below 20 MeV).  As for additional physics, COALESCE card was activated to ensure light cluster formation takes place, which is also included in FLUKA's PEANUT model (Pre-Equilibrium and Evaporation Nuclear Model).  This model is particularly useful for nuclear inelastic interactions for neutron energies above 20 MeV. Below this threshold, nuclear elastic scatterings are based on Ranft model (see Ref.~\cite{Ranft1972}) and FLUKA follows group-wise approach for treatment of neutrons. This means at each energy group, reaction cross-sections are taken from evaluated nuclear databases such as ENDF, JEFF, JENDL are averaged over neutron fluence. Cross-sections calculations are hence concluded by equation \eqref{eq:xs}\cite{fluenceformulaSisti}
\begin{equation}
\sigma_i = \frac{\int_{E_{i, \text{low}}}^{E_{i, \text{high}} } \sigma(E) \Phi(E) d E}{\int_{E_{i, \text{low}}}^{E_{i, \text{high}}} \Phi(E) d E}\label{eq:xs}
\end{equation}
Here, $\Phi(E)$ represents the particle fluence (number of incident particles per unit area, typically in $cm^{-2}$, $\sigma(E)$ represents the cross-section and \textit{E} is the energy.  Due to the projectile particle being a proton beam having energy in the order of few MeV, the coalescence process could result in the formation of light nuclei like, ${}^2\mathrm{H},  {}^3\mathrm{H}, {}^3\mathrm{He}, {}^4\mathrm{He}$ which can indirectly contribute to neutron and gamma yield via further interactions~\cite{coalesce}.
To count neutrons more accurately, decays to neutrons shall also be taken into consideration, hence DECAYS card was also activated. Another neutron source is the nuclear evaporation process. Neutron emission from excited nuclei was activated using EVAPORAT card.
All calculations were done using built-in scoring by utilization of various estimators.

 \section{Results}
 \subsection{Target Orientation and Angular Distribution of Neutrons}
Ideal target orientation was determined by analyzing double differential fluence of neutrons using USRBDX detector card. Due to the quasi-logarithmic group energy structure of neutrons, logarithmic energy binning was imposed with 200 energy bins to study the wide energy spectrum. For the solid angle binning, we set up 15 angular bins, starting from 0 to $\pi / 2$. This angle is with respect to surface normal (i.e. target). As the target itself is positioned 45$^\circ$ angle with respect to origin, another 45$^\circ$ angle with respect to surface normal is equivalent to 90$^\circ$ angle with respect to beam direction(i.e. polar angle $\theta$ where \textit{z} axis is the beam direction).  
\begin{figure}[h!]
    \centering
    \begin{minipage}{0.6\linewidth} 
        \centering
        \includegraphics[width=\linewidth]{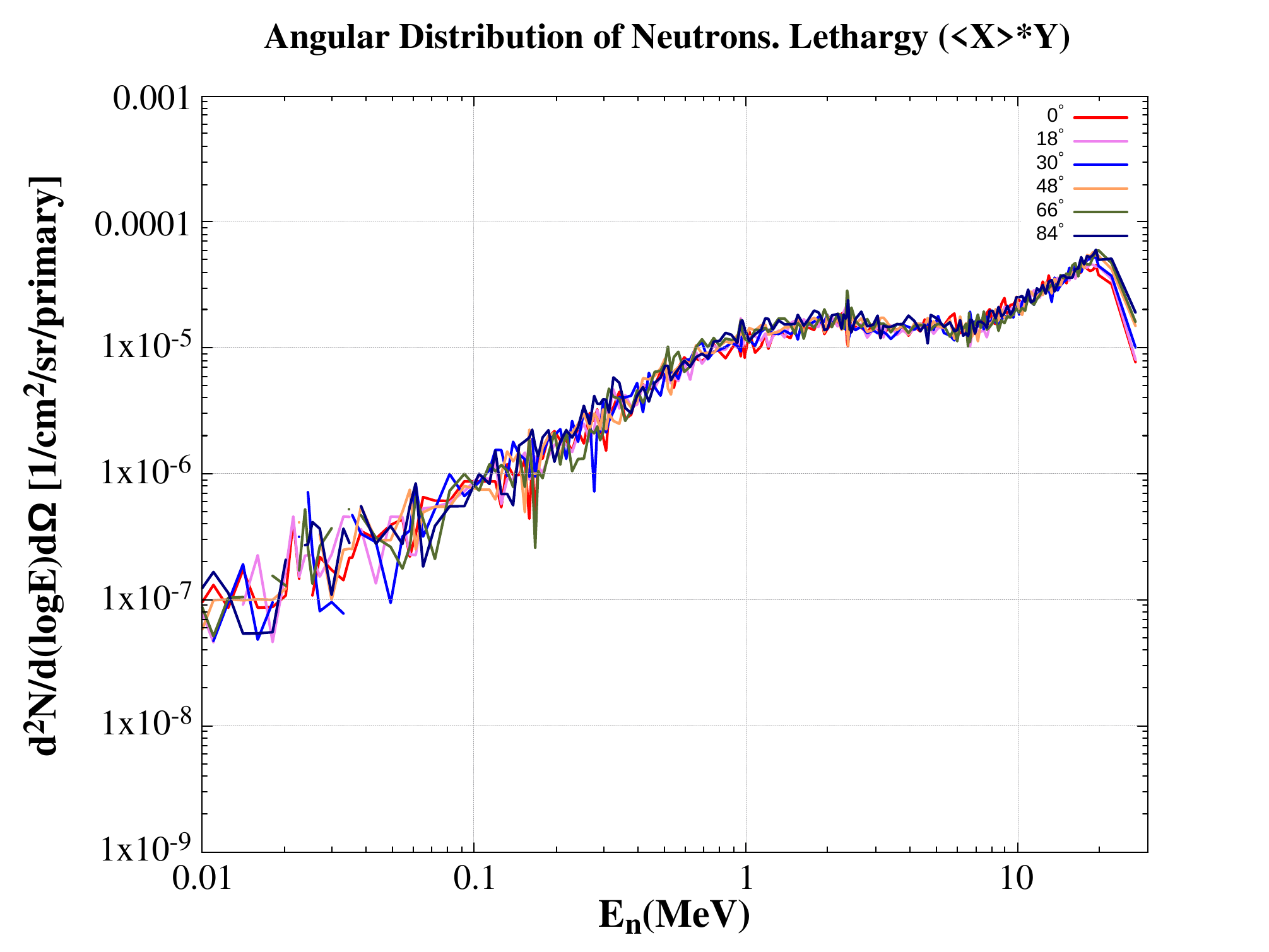}
    \end{minipage}
    \hfill
    \begin{minipage}{0.6\linewidth}
        \centering
        \includegraphics[width=\linewidth]{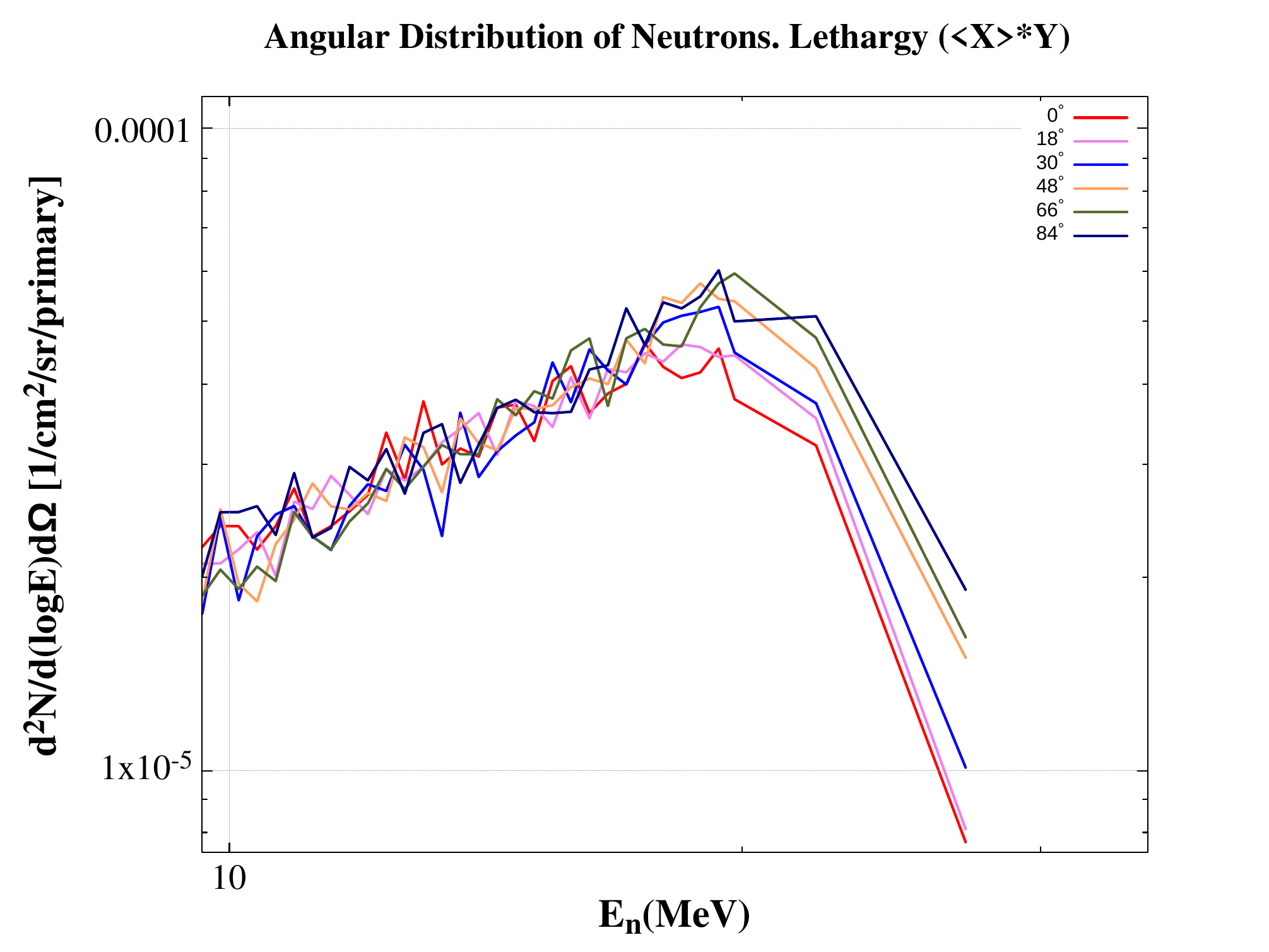}
    \end{minipage}\caption{Azimuthal ($\phi$) distribution of resulting neutrons~\cite{göver_2025}.}\label{fig:angular_neutron_dist}
\end{figure}
Next, we calculated the angular distribution of neutrons with respect to azimuthal angle. In  Figure \ref{fig:angular_neutron_dist},  angular distribution of neutrons was registered. Neutron scattering events within a target material induce a deviation in the neutron fluence trajectory, shifting it towards a more orthogonal direction. The figure on the right shows the high tendency for higher angles.  For neutrons of higher energies, scattering emerges as the primary mechanism governing the channeling (or collimation) of the neutron beam. At lower energies, we observe the exact opposite.  Lower energy neutrons exhibit a greater tendency to travel along paths parallel to the coordinate axes, hence motion along the directions defined by the orthogonal coordinate axes is more prevalent at lower neutron energies.

\begin{figure}[h]
    \centering
    \begin{minipage}{0.49\textwidth} 
        \centering
        \includegraphics[width=\linewidth]{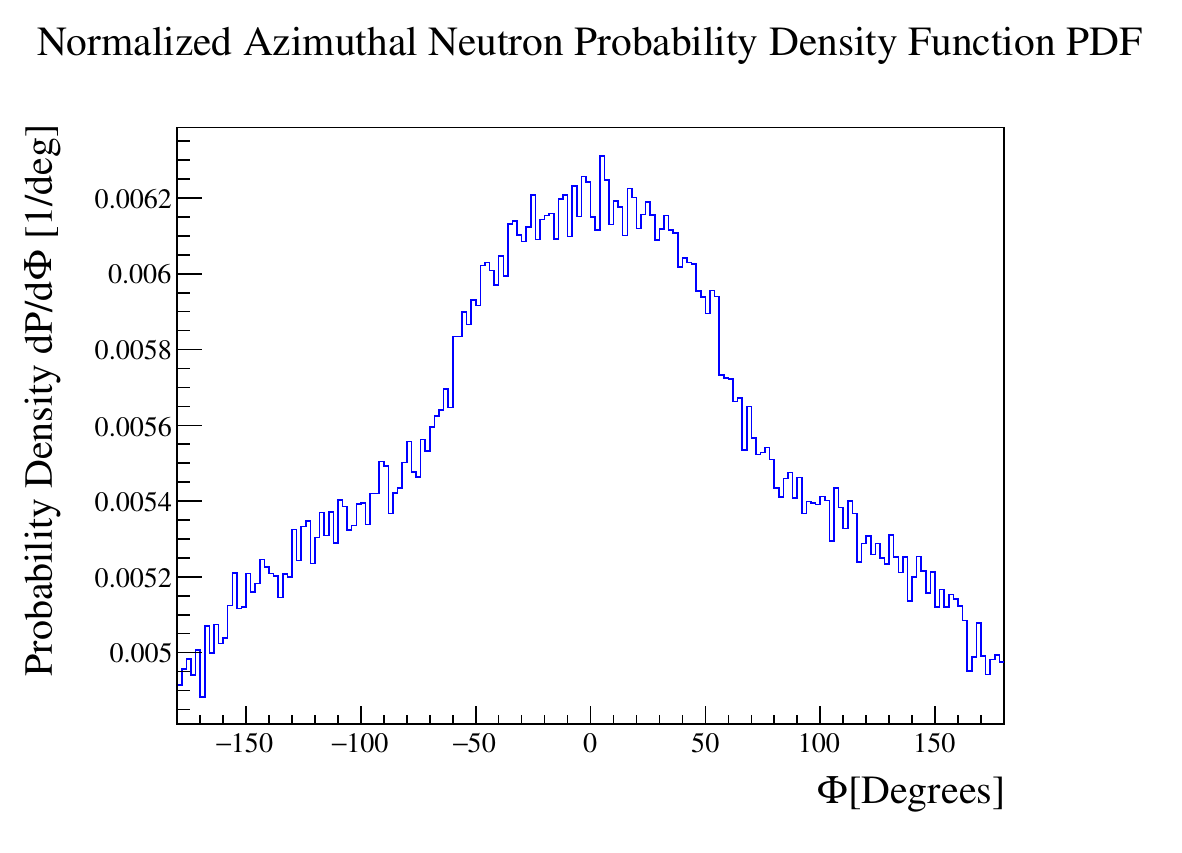}
    \end{minipage}
    \hfill
    \begin{minipage}{0.49\textwidth}
        \centering
        \includegraphics[width=\linewidth]{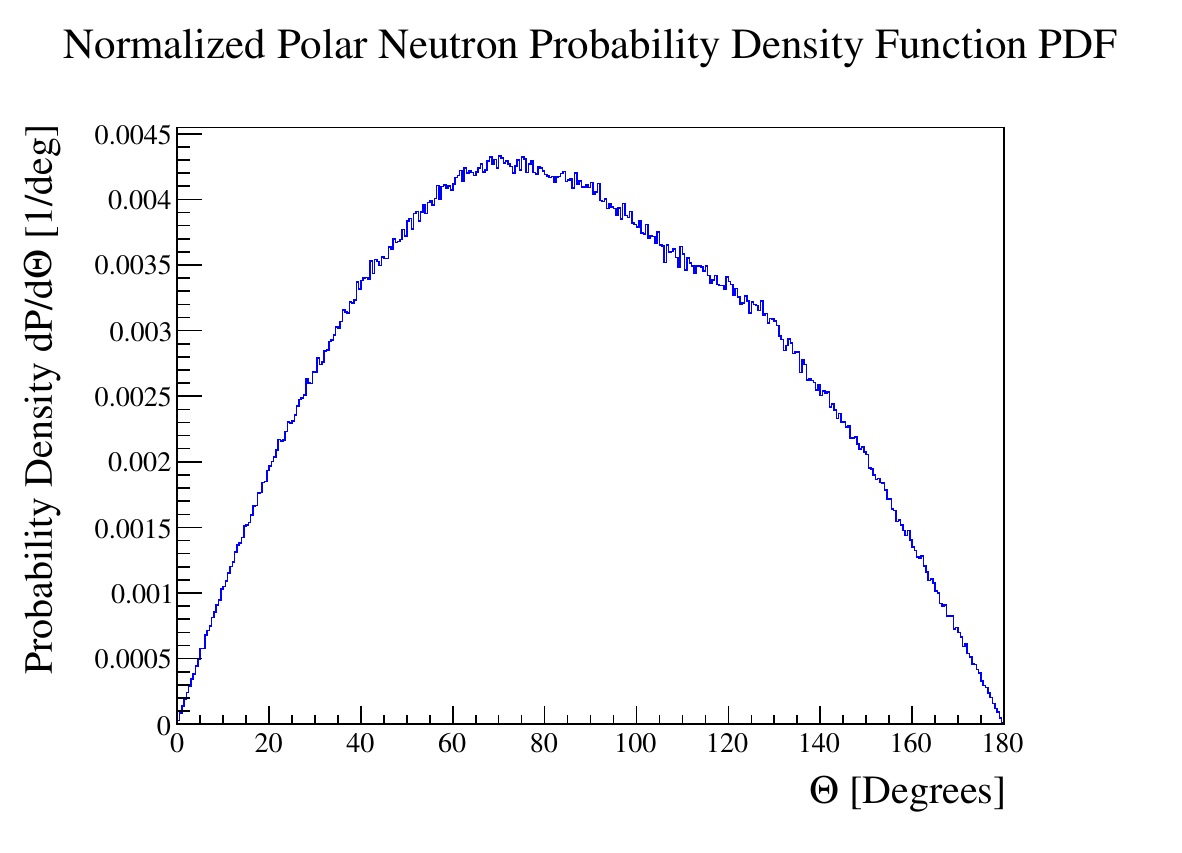}
       
    \end{minipage} \caption{Spherical angular distribution of neutrons emerging from beryllium target. \cite{göver_2025}.}\label{fig:theta}
\end{figure}

Using Geant4, a thin spherical shell was  used to encapsulate beryllium target to register the angular distribution of outgoing neutrons.  A peak in neutron count around $\theta$  $\sim 70^\circ$ was observed. The probability density function of $\Theta$ depicted in Figure \ref{fig:theta} (right) demonstrates a leftward spectral shift within the visible range. This shift, interpreted as a consequence of the initial target orientation, indicates a contribution of 45$^\circ$ target tilt to the observed beam orientation.
\subsection{Neutron Fluence and Moderation}
\begin{figure}
    \centering
    \includegraphics[width=0.9\linewidth]{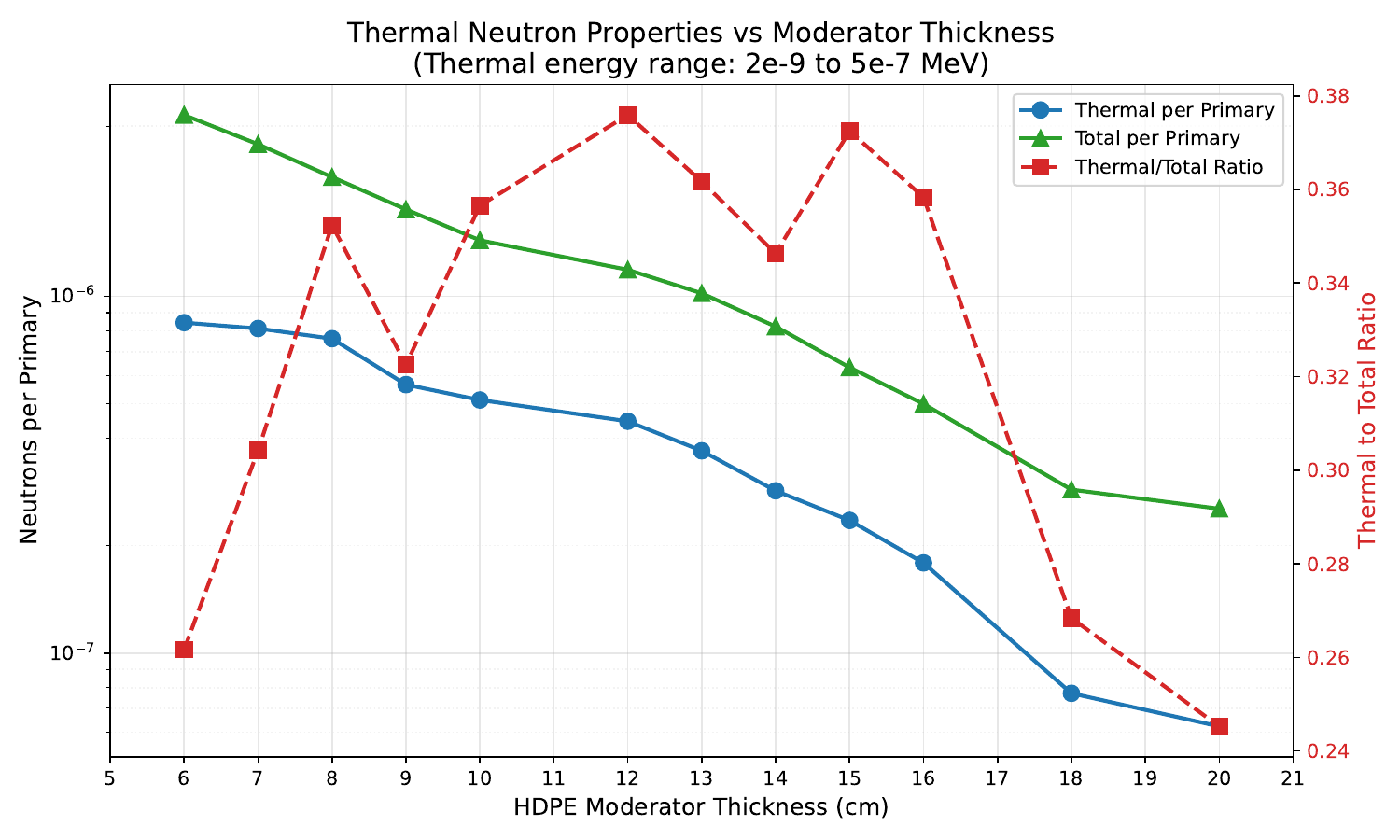}
    \caption{Various thicknesses of high density polyethylene was simulated to find thermal ratio and primary response at P(70).}
    \label{fig:variousthck}
\end{figure}

\begin{figure}[h]
    \centering
    \includegraphics[width=0.6\textwidth]{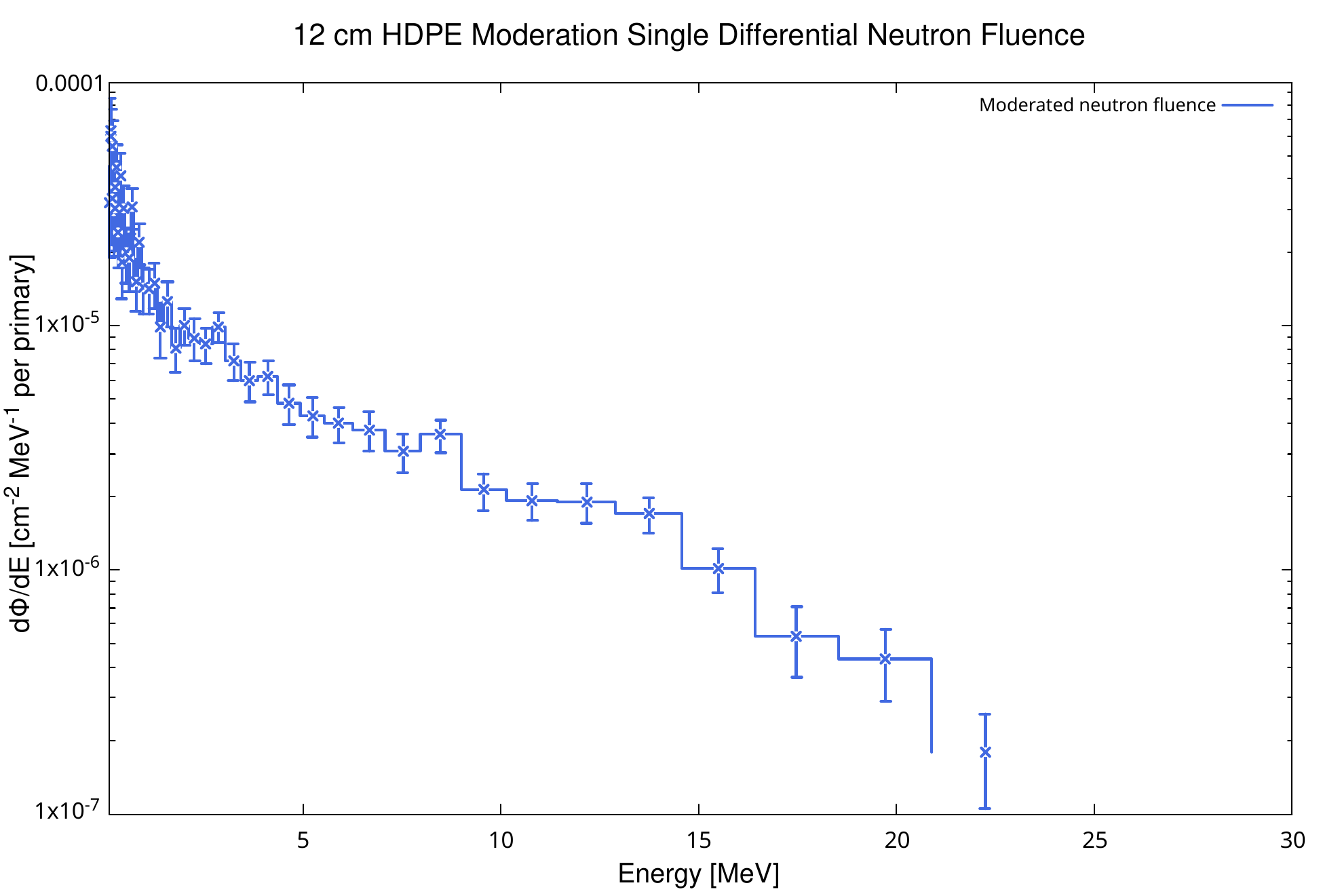}
        \caption{Single differential moderated neutron fluence at P(70).}
    \label{target_usrtrack}
\end{figure}

In order to compare how neutron output fluence behaves under various materials and moderation stages, initial resulting neutron count from proton-beryllium reaction was calculated. For the probe point, the volume is 4.19 $\text{cm}^3$ which is used by USRTRACK to normalize fluence (integrated over solid angle $\Omega$). Figure \ref{target_usrtrack}  shows the resulting energy spectrum and fluence of neutrons inside the target.
Optimal polyethylene thickness was determined to achieve the highest thermal to total neutron ratio. Various thicknesses were tested in Figure \ref{fig:variousthck}. For 12.0 cm, a thermal ratio of 37.6\% was achieved. It was also tested to see if encapsulating target with HDPE block would yield a better neutron count. Table \ref{tab:neutron_stats} compares the configuration with an HDPE slab positioned in the probe direction to the configuration with an HDPE box fully enclosing the target.

\begin{table}[ht]
    \centering
    \caption{Neutron Statistics at P(70) for Different Moderator Configurations. 1E8 primary proton.}
    \label{tab:neutron_stats}
    \begin{tabular}{l c c c c c}
        \toprule
        Case & \multicolumn{2}{c}{Thermal Neutrons\textsuperscript{a}} & \multicolumn{2}{c}{Total Neutrons} & Thermal/Total \\
        \cmidrule(lr){2-3} \cmidrule(lr){4-5}
        & Count ($\pm \sigma$) & Per Primary & Count & Per Primary & Ratio \\
        \midrule
        12 cm (Slab) & $13.33 \pm 1.75$ & $1.33 \times 10^{-7}$ & 62.79 & $6.28 \times 10^{-7}$ & 0.2123 \\
        12 cm (Encap.)     & $44.65 \pm 4.01$ & $4.46 \times 10^{-7}$ & 118.78 & $1.88 \times 10^{-6}$ & 0.3759 \\
        \bottomrule
        \addlinespace
        \multicolumn{6}{l}{\textsuperscript{a} \footnotesize Energy range: $[2.00 \times 10^{-9}, 5.00 \times 10^{-7}]$ MeV}
    \end{tabular}
\end{table}

\begin{figure}[h]
    \centering
    \includegraphics[width=0.55\linewidth]{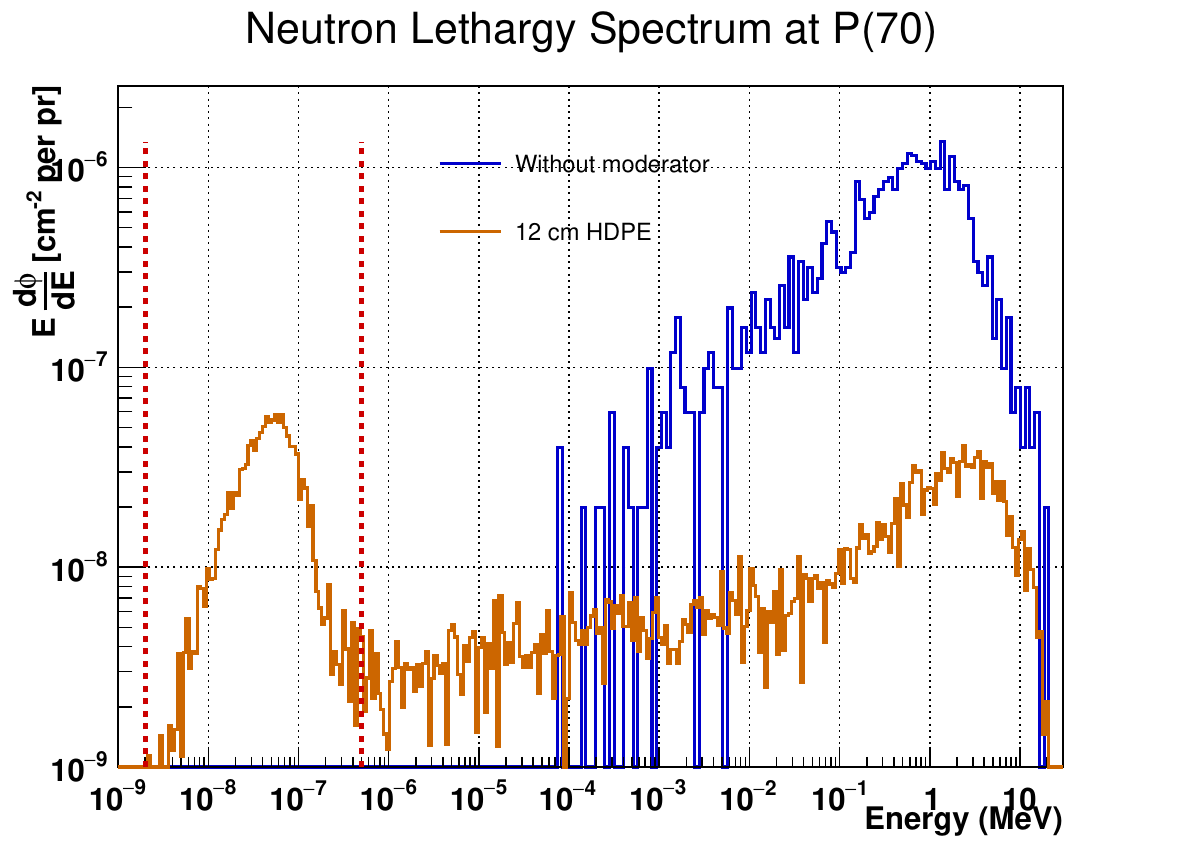}
    \caption{Moderated neutron flux at P(70) due to 12 cm thick polyethylene. The thermal energy band is shown between dotted red lines.}
    \label{fig:12cmhdpeModeration}
\end{figure}

Figure \ref{fig:12cmhdpeModeration} shows the Geant4 results of moderation done by polyethylene block of 12 cm, at point P(70), positioned 70 cm away from the target. The observed increase in thermal neutron concentration, as evidenced by the heightened intensity of closer peaks in lower energy bands, suggests a moderating effect induced by the presence of polyethylene. In the thermal energy band [2.00$\times 10^{-9}$ MeV, 5.00$\times10^{-7}$ MeV], thermal neutron yield given per primary was found as 4.46$\times10^{-7}$.

\begin{figure}[h]
    \centering
    \includegraphics[width=0.6\linewidth]{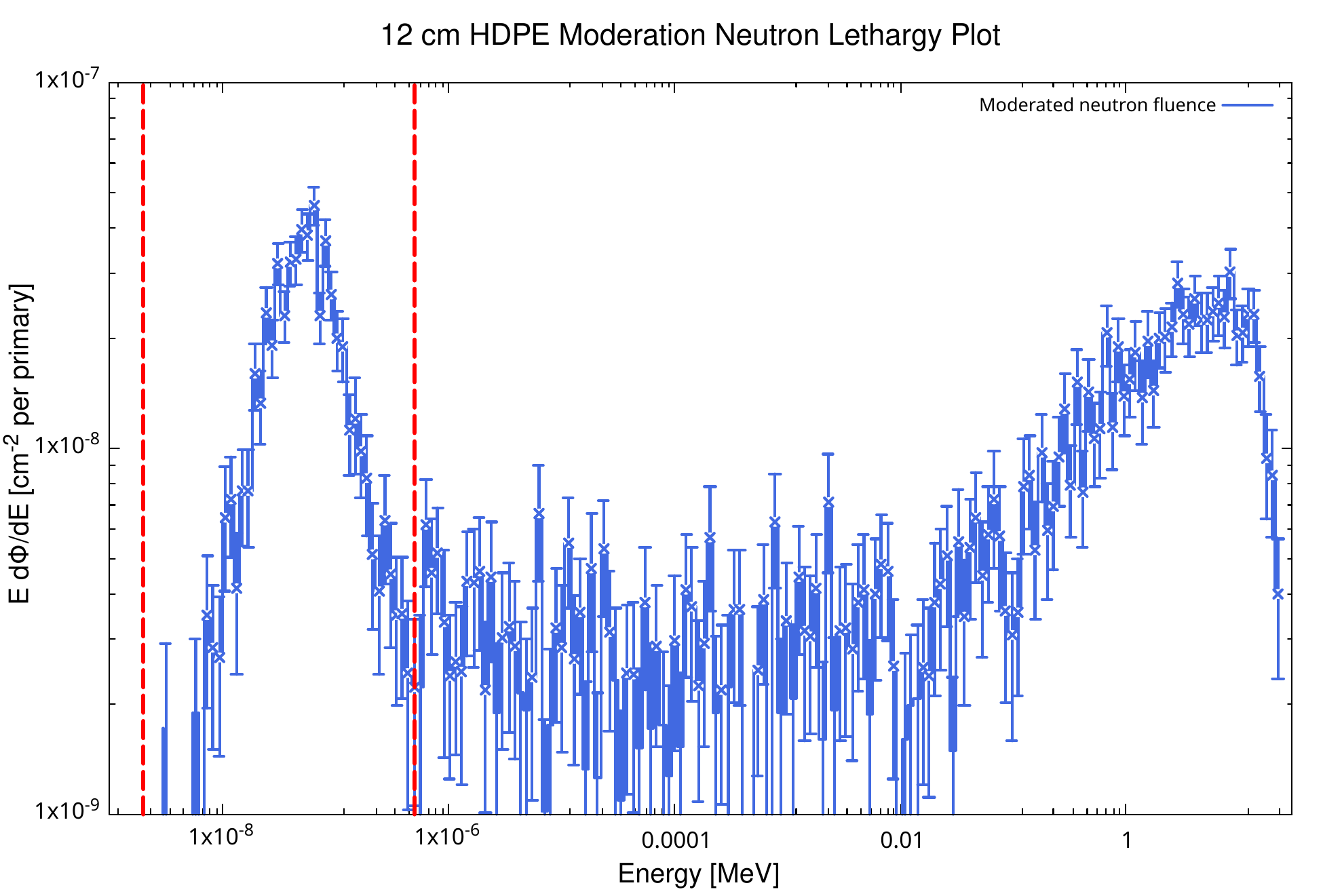}
    \caption{Lethargy plot of moderated neutron fluence at P(70). Thermal band is shown between dotted red lines.}
    \label{fig:targetusrtrack_lethargy}
\end{figure}

For comparison, FLUKA results were also given in Figure \ref{target_usrtrack} and in Figure \ref{fig:targetusrtrack_lethargy}. There is a considerable amount of neutrons present at significantly lower energies, extending to $\sim 10^{-8}$ MeV. 
At  lower energies (< 1 MeV), FLUKA and Geant4 agree on neutron spectral fluence. At higher energies, the results tend to deviate from each other. 

\begin{figure}[h!]
\centering
\includegraphics[width=.7\textwidth]{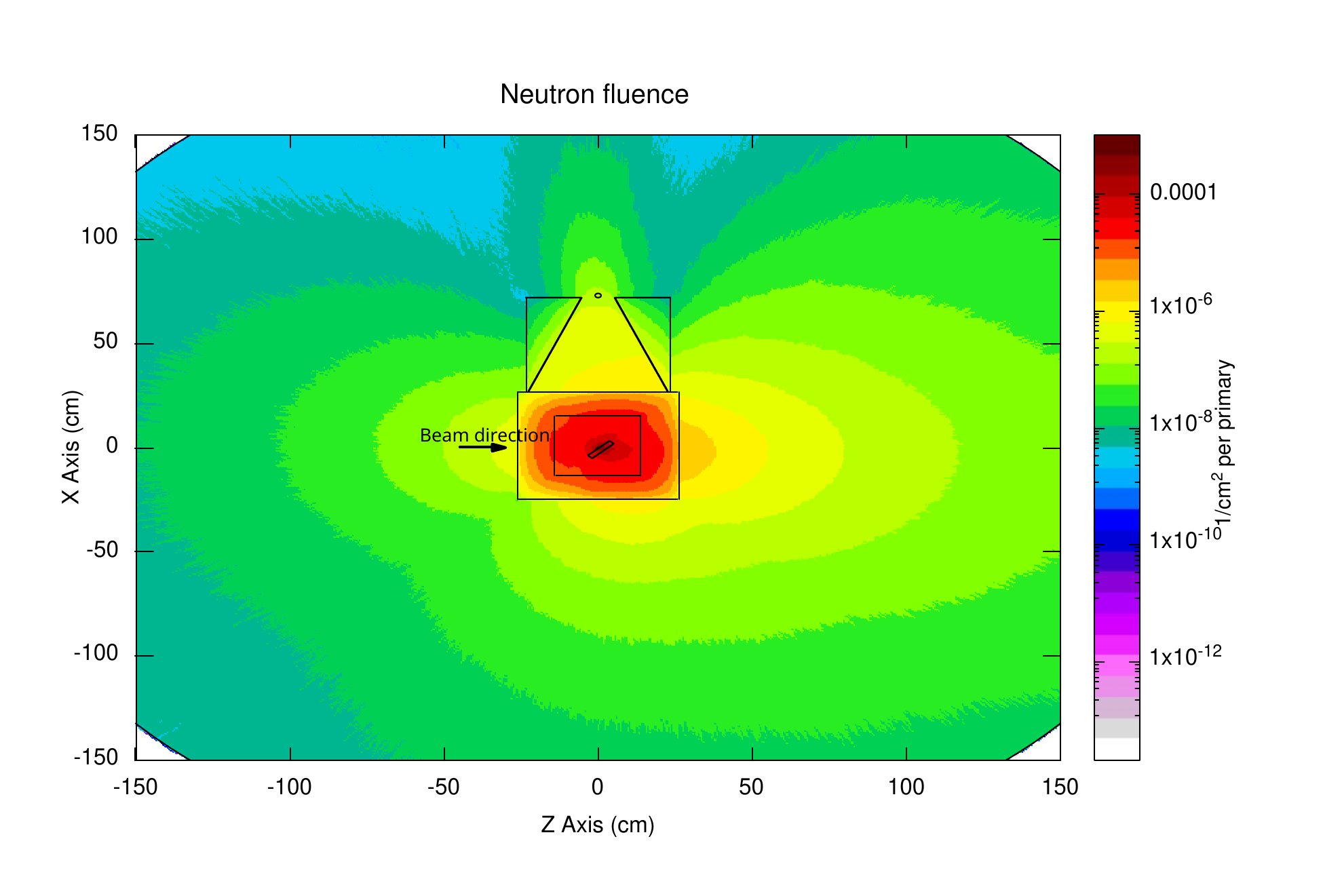}
\caption{Neutron fluence across the Y plane given in per primary.\label{usrbin:notrfluence}}
\end{figure}

\begin{figure}[h]
    \centering
    \includegraphics[width=0.7\linewidth]{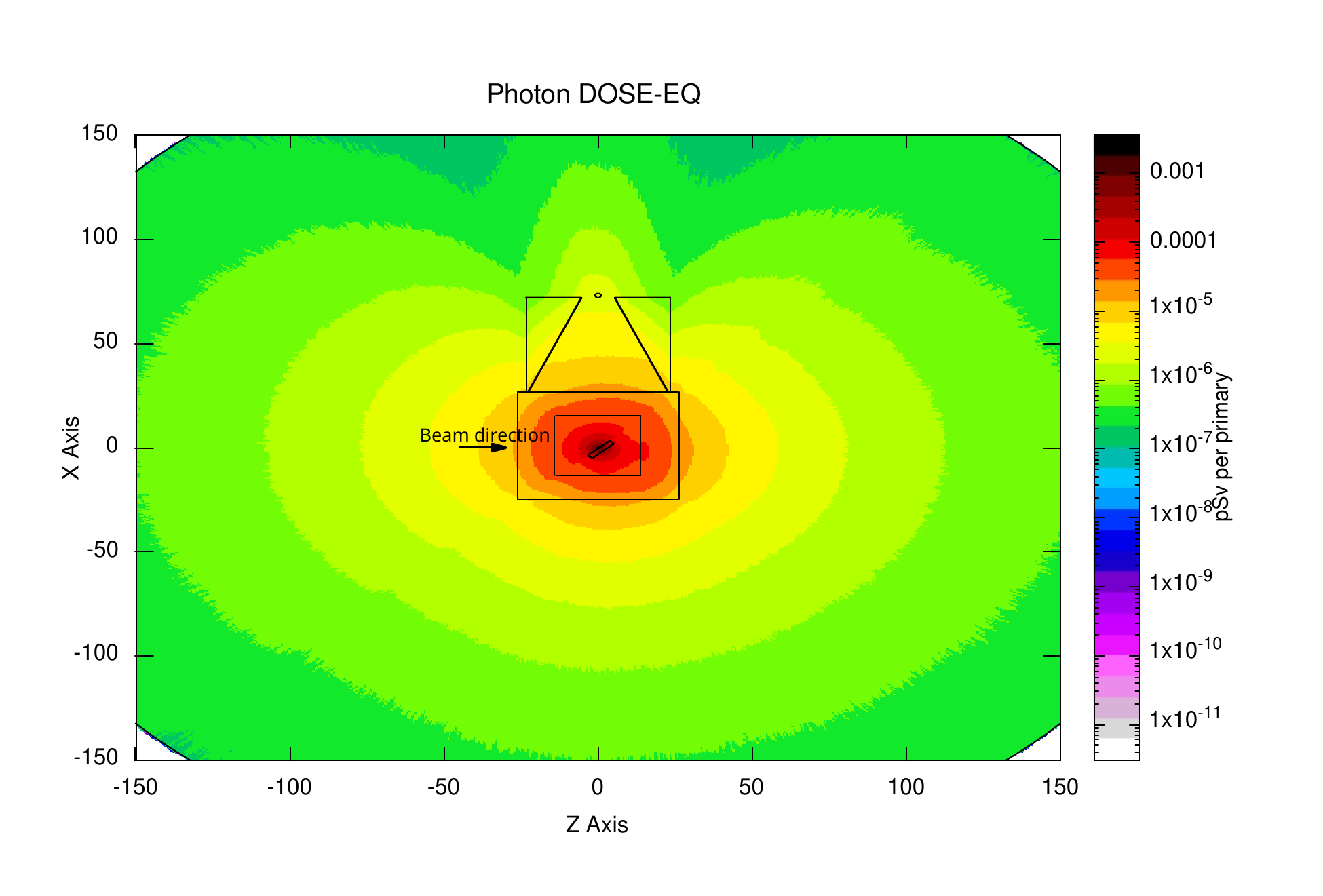}
    \caption{Prompt photon dose equivalent across the Y plane given in pSv per primary (no shielding).}
    \label{fig:photon_Doseeq}
\end{figure}

\begin{figure}[h]
    \centering
    \includegraphics[width=0.7\linewidth]{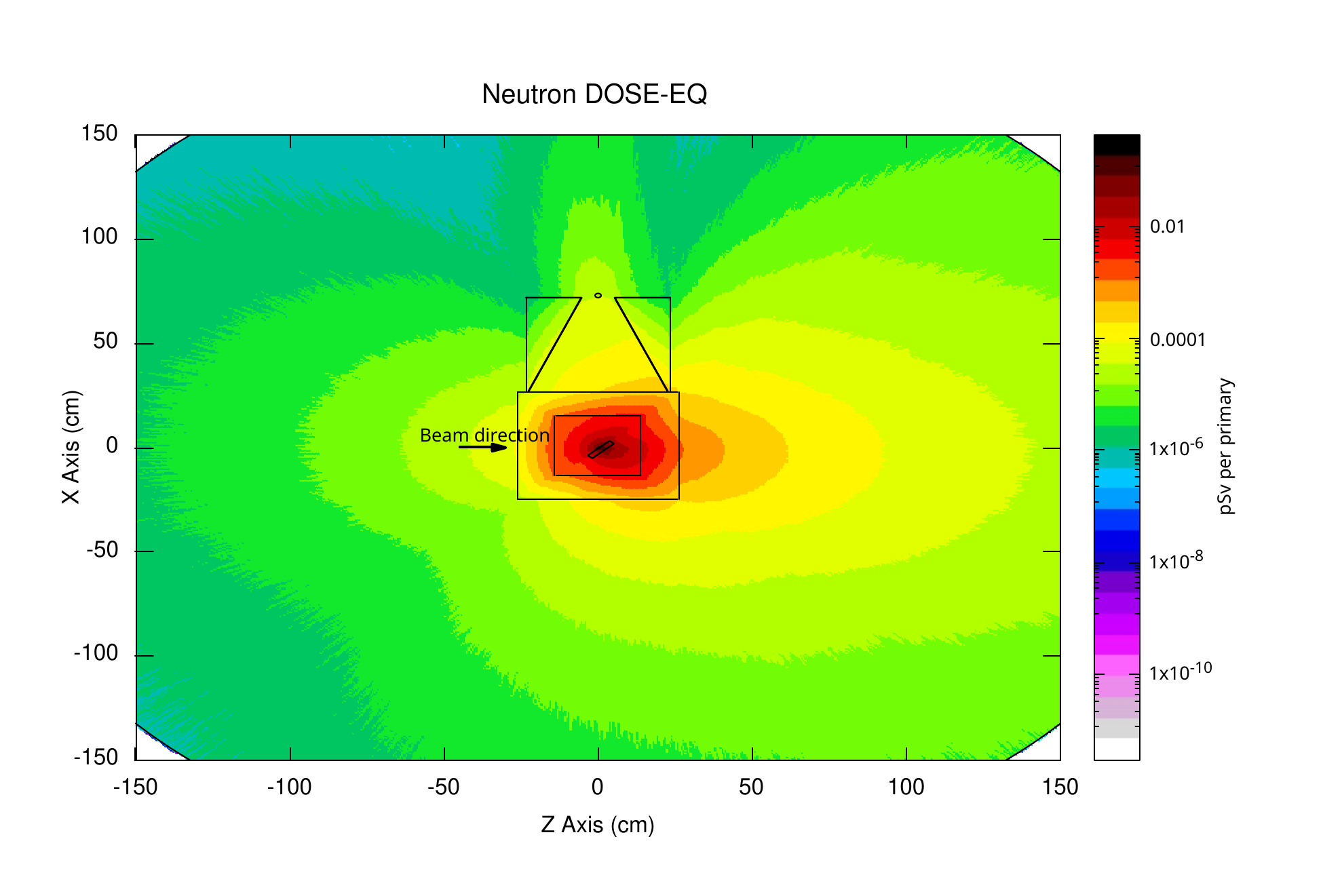}
    \caption{Prompt neutron dose equivalent across the Y plane given in pSv per primary (no shielding).}
    \label{fig:neutron_Doseeq}
\end{figure}

The general neutron fluence across the irradiation chamber was calculated using USRBIN, which can be seen in Figure \ref{usrbin:notrfluence}. The neutron funnel successfully channels the neutrons in the desired direction.
To calculate the dose inside the chamber, where neutron and gamma radiation contribute individually, USRBIN was utilized as well. Figure \ref{fig:photon_Doseeq} and \ref{fig:neutron_Doseeq} show the photon and neutron equivalent dose inside the chamber. The results were given in the units of pSv/primary.

\subsection{Neutron Beam Distribution Characteristics}
\begin{figure}[h!]
    \centering
    \includegraphics[width=1\textwidth]{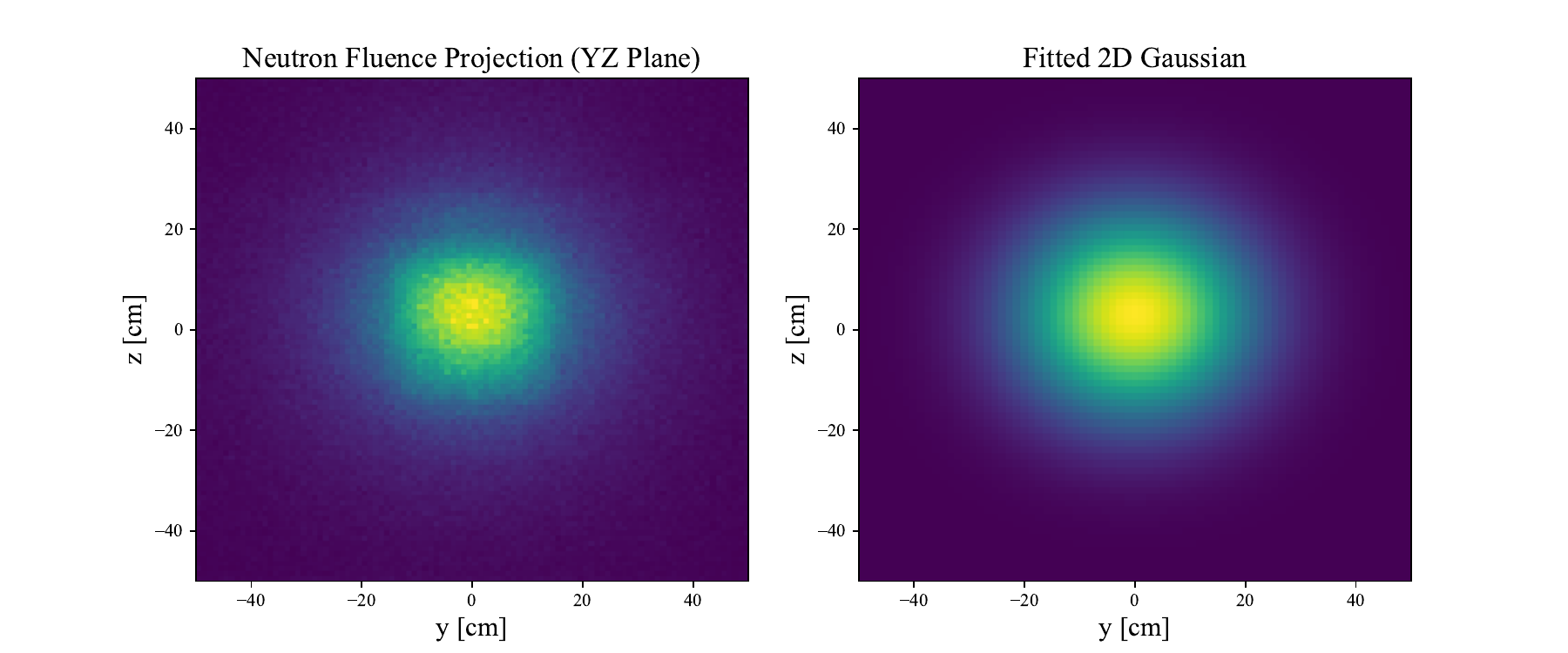}
    \caption{Neutron Fluence Projection and Fitted 2D Gaussian Distribution}
    \label{fig:gaussl}
\end{figure}

\begin{figure}[h]
    \centering
    \includegraphics[width=0.6\textwidth]{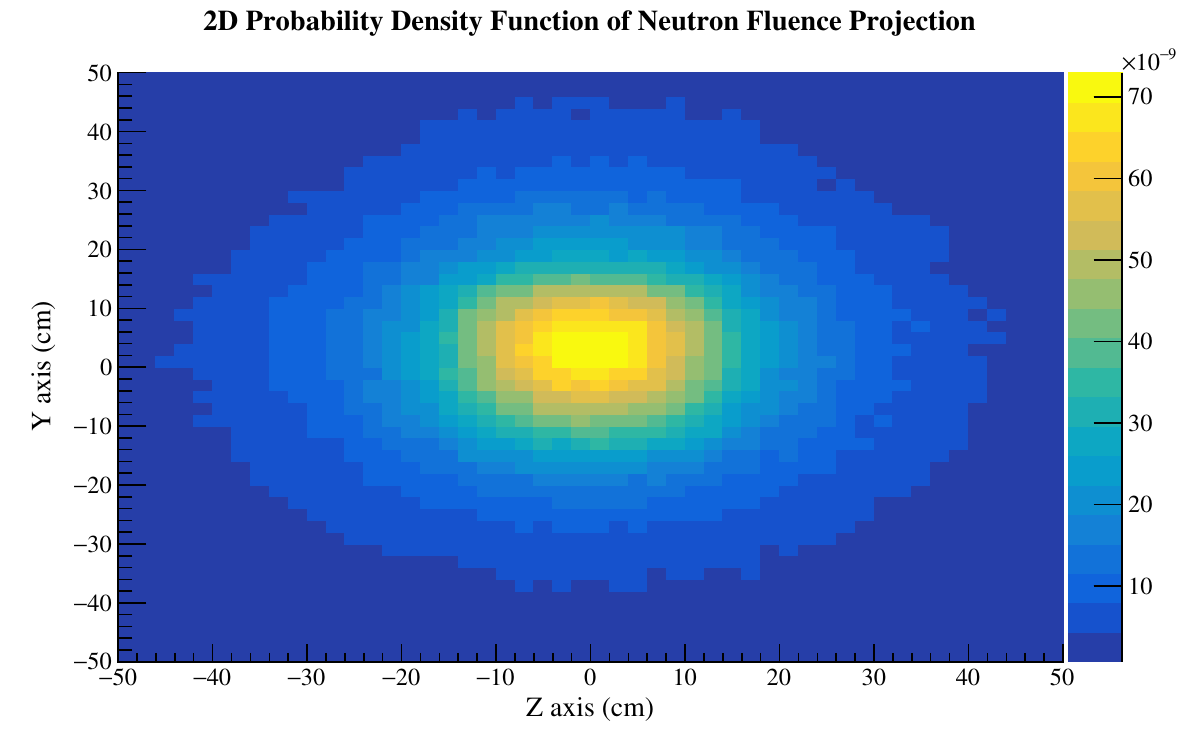}
    \caption{Gaussian beam profile given in probability distribution across YZ plane (COLZ representation).}
    \label{fig:colz}
\end{figure}

\begin{figure}[htbp!]
    \centering
    \includegraphics[width=0.7\textwidth]{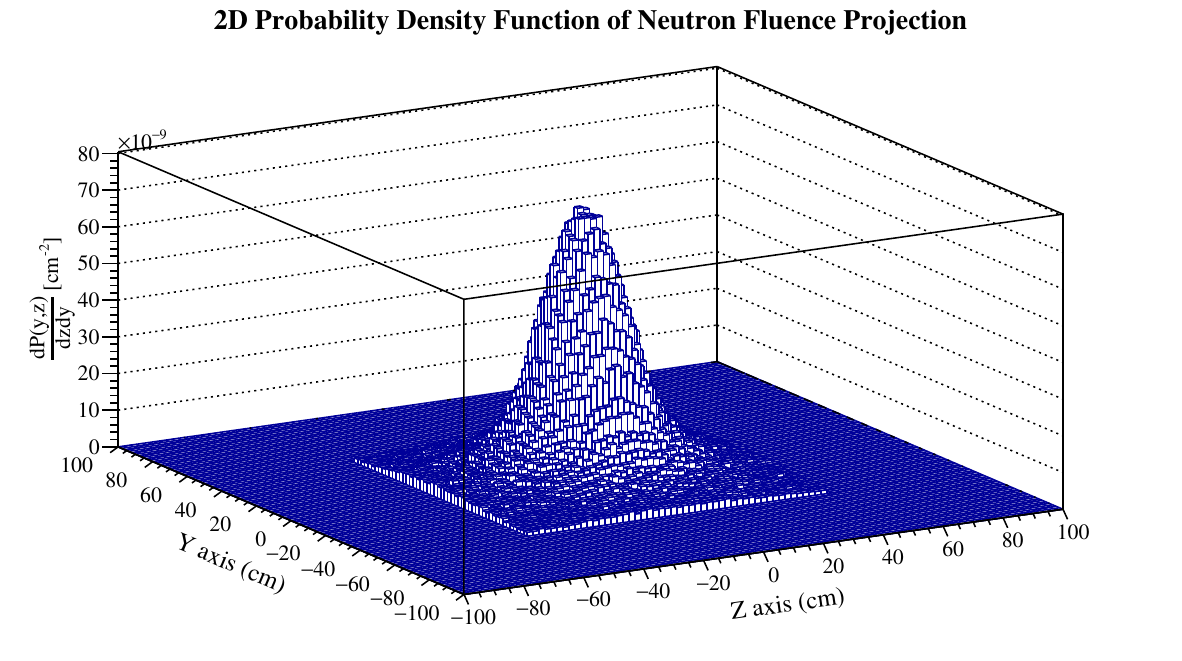}
    \caption{Gaussian beam profile given in probability distribution across YZ plane (LEGO representation).}
    \label{fig:lego}
\end{figure}

Figure \ref{fig:gaussl} shows the beam profile projected onto YZ plane. Constructed neutron beam was found to follow a Gaussian profile. To determine the fitted 2D Gaussian of the neutron beam , we used\cite{gauss} 
\begin{equation}
f(z, y)=A \exp \left(-\left(\frac{\left(z-z_0\right)^2}{2 \sigma_Z^2}+\frac{\left(y-y_0\right)^2}{2 \sigma_Y^2}\right)\right)
\end{equation}
where the fitted parameters were extracted from ROOT file. The calculated Gaussian parameters were as follows: A=749.252, $y_0$=0.009 cm, $z_0$=2.942 cm with standard deviations $\sigma_{y}$=14.874 cm, $\sigma_{z}$=14.220 cm. For comparison, the FWHM of incoming proton beam was 1.5 cm with $\sigma = 0.636$ cm. Figure \ref{fig:colz} and \ref{fig:lego} shows the LEGO plot of beam profile in 2D and 3D respectively. The neutron beam projection spans a large area, which extends to $\sim 100 \text{ cm}$ in each direction. 

\section{Conclusion}
This work presented the preliminary trial run of  a proposed proton-beryllium irradiation experiment.
The numerical data obtained from FLUKA using the PRECISION physics as default setting is in more parallel with Geant4 when the physics list specified in Section \ref{physics_list} is registered. We found at higher neutron energies, FLUKA seems to score higher neutron fluence than Geant4, while at lower energies these two simulation toolkits agree on fluence estimation.

Beryllium is a compelling target material for accelerator-driven neutron sources. The significant flexibility in tailoring the neutron energy spectrum (which can be done by varying the incident particle energy in the accelerator and the moderators) enables the feasibility of a wide range of neutron-based studies. The angular aspect provides an additional degree of freedom for shaping the neutron beam.
The low atomic number of beryllium (Z=4) and relatively low neutron separation energy (1.67 MeV) enables efficient neutron production through reactions such as $^{9}\mathrm{Be}(p,n)^{9}\mathrm{B}$ and $^{9}\mathrm{Be}(d,n)^{10}\mathrm{B}$.  Its relatively high melting point and thermal conductivity makes it convenient to implement liquid water as primary cooling mechanism, which would be an efficient and cheap alternative to cryogenic cooling.  However, beryllium should be handled with caution, as it is highly toxic, volatile under high-intensity irradiation and requires strict safety protocols.

\appendix

\acknowledgments

DV is grateful to The Scientific and Technological Research Council of T\"urkiye for their support through the grants 123C484 and 125F379. DV acknowledges further support from the European Union’s Horizon Europe Research and Innovation Programme under the Marie Skłodowska-Curie
Actions (MSCA) COFUND Programme with the grant
number 101081645 and The Scientific and Technological Research Council of T\"urkiye with the grant number 123C213. We extend our sincere gratitude to Middle East Technical University for their generous support.



\bibliographystyle{ieeetr}
\bibliography{biblio}

\end{document}